# Temperature effect on the cardiac ryanodine receptor gating and conductance: mathematical modelling

A.S. Moskvin[*], B.I. Iaparov[*], A.M. Ryvkin[* **], O.E. Solovyova[* **]

[*] Ural Federal University, 51, Lenin str., 620083, Ekaterinburg

[**]Institute of Immunology and Physiology UB RAS, 106, Pervomayskaya str., 620049, Ekaterinburg,

E-mail: bogdan.iaparov@urfu.ru

The temperature effect on the cardiac ryanodine receptor (RyR) function has been studied within the electron-conformational (EC) model. It's shown that simple EC model with the Arrhenius like temperature dependence of "internal" and "external" frictions and a specific thermosensitivity of the tunnelling "open ↔ close" transitions can provide both qualitative and quantitative description of the temperature effects for isolated RyRs. The potential of the model was illustrated by explaining the experimental data on the temperature dependence of sheep's isolated cardiac RyR gating and conductance (R. Sitsapesan et al., J Physiol **434**,469(1991)).

## I. INTRODUCTION

Excitation-contraction coupling (ECC) is the process by which excitation of the myocyte leads to cardiac contraction [1]. ECC is very sensitive to a temperature change. Hypothermia prolongs action potential duration (APD), and increases contraction and $Ca^{2+}$ transient amplitudes [2]. Single-channel studies of L-type calcium release channel[3,4,5], sodium-calcium exchanger[6] and cardiac ryanodine-sensitive calcium release channel(RyR channel) [7] regulation by temperature show, that one of the main reasons of ECC dependence on temperature are changes in cardiomyocytes' ion channels with temperature.

A main way to characterize temperature sensitivity is the $Q_{10}$ value which is equal to fractional increase of any variable upon a 10°C increase in temperature. For most ion channels $Q_{10}$ varies between 2 and 5. Among the exceptions are a group of temperature-sensitive transient receptor potential (TRP) channels[8], which are in sensory nerves and provide temperature sensitivity, $Q_{10}$ for these channels varies between 2 and 100 [9].

In this work we dwell on a mathematical modeling of temperature effects on ligand-gated RyR channel. This largest known ion channel with molecular mass ~ 2.2 MDa [10] plays key role in intracellular calcium dynamics in cardiomyocytes, RyR channel's dysfunction leads to arrhythmias and heart failure [11].

Single-channel experiments performed on cardiac sheep RyR channels[7] show, that cooling from 23 to 5 °C leads to qualitative changes of current fluctuations with increase of open probability $P_{open}$($Q_{10}$ = 2.5) by increasing mean open time $<t_{open}>$($Q_{10}$ = 5.7) and not significantly changing mean closed time $<t_{close}>$ ($Q_{10}$ = 1.1). From the viewpoint of the condensed matter physics it seems to be absolutely inexplicable how the ~3% change in absolute temperature can produce many-fold effect in the channel activity.

Here, in this paper, we show, that an earlier proposed electron-conformational model (ECM) of RyR channels[12-14] successfully reproduces the experimentally found temperature behavior of the RyR channels[7] and unlike purely phenomenological models [9,15,16], uncovers some universal features of the physiological thermosensitivity. Preliminary results of this work were published as letter [17].

## II. ELECTRON-CONFORMATIONAL MODEL

RyR channel is the biggest known ion channel with a tetrameric structure and molecular mass of ~ 2.2 MDa [19]. As many other ion channels it has many internal electronic and conformational degrees of freedom. Channel gating is typically modeled by aggregated Markov chains [18], which are, in fact, a generalization of the simplest «hole in the wall» model with a set of discrete open and closed states with each state connected to its neighbors through unimolecular rate constants. Numbers of states in different classes (open, closed) are selected according to maximum likelihood method (for example [19]).

Obviously these models, despite their popularity, cannot provide a reasonable molecular mechanistic insight on the origin of specific properties of RyR channels gating.

In papers [12-14] an electron-conformational model (ECM) of an isolated RyR channel and RyR channel clusters was proposed, which is based on a generalization of a widely known theory of photo-induced phase transitions [20,21]. ECM is able to describe main features of the behavior of isolated and interacted ion channels in cardiomyocytes and pacemaker cells. Within

ECM the RyR channel is described likely an elastic rubber tube with a varying cross-section governed by a conformational variable Q and a light "electronic" plug switched due to $Ca^{2+}$-RyR binding/unbinding. This electronic plug interacts with the conformational variable and acts as a trigger to stimulate its change and related channel conductivity. In other words, a large variety of the RyR degrees of freedom one reduces to only two: a fast and a slow one conventionally termed as the electronic and conformational one, respectively. The simplest ECM version implies only two channel's electronic states: the «open» and «closed» ones with conformational variable Q assumed to be continuous. A scheme of the ECM RyR channel gating is illustrated in Fig.1, where the typical states are denoted by small (o,c) and capital (O,C) letters. Changing electronic or conformational states regulates main RyR channel's function and defines if the channel is open and $Ca^{2+}$ ions can pass through ion channel. Conformational variable Q specifies the RyR channel "crosssection" or, more precisely, a permeability for $Ca^{2+}$, that is a conductance, while the dichotomic electronic variable determines its opening and closure. $Ca^{2+}$ flow through an isolated ion channel is defined as follows:

$$J_{RyR}(Q) = k_{rel} D(Q)(Ca_{jSR} - Ca_{SS}) \quad , \qquad (1)$$

where $k_{rel}$ is the release flux rate constant, and the functional dependence $D(Q)$ is related to the level of the channel opening. Here we assume a power dependence on the conformational variable:

$$D(Q) = D_0 Q^\beta \quad , \qquad (2)$$

when $Q>0$ and channel is electronically open and $D=0$ in other cases.

The key element of ECM is a so called «energetic» approach, in other words, namely the energy is supposed to be a main characteristic of the RyR channel's state. We introduce a simple effective Hamiltonian, or the energy operator for an isolated RyR channel as follows:

$$\hat{H} = -\Delta \hat{s}_z - h \hat{s}_x - pQ + \frac{KQ^2}{2} + aQ\hat{s}_z \quad , \qquad (3)$$

where $s_x$, $s_z$ are Pauli matrices, first two terms describe purely electronic effects: first term describes the bare energy splitting of "up" and "down" (electronically "open" and "closed") states with an energy gap $\Delta$, the second term describes quantum effects of transition between two electronic states. Next two terms are the linear and quadratic contributions to the channel's "elastic" energy, moreover, the linear term corresponds formally to an energy of external forces, which are described by an «effective pressure» *p*, *K* is the channel's effective elastic constant. Last term describes electron-conformational interaction with a coupling constant *a*. We make use of the dimensionless conformational variable Q; therefore all of the model parameters (Δ, h, p, K, a) are assigned energy units.

Two eigenvalues of Hamiltonian define two branches of the adiabatic conformational potential (CP):

$$E(\mu, Q) = \frac{KQ^2}{2} - pQ \pm \frac{1}{2}\sqrt{(\Delta - aQ)^2 + h^2}, \qquad (4)$$

which implies appearance under certain conditions of the bistability effect, or two minimas of CP, located left and right from the CP intersection point and attributed to electronically and conformationally closed (cC) and open (oO) channel states, respectively. For simplicity we take in the paper the case when *h*=0, that implies so called diabatic CP (see Fig.1).

Unfortunately, there's neither experimental nor microscopic calculations data, which could provide independent trusted information about ECM parameters and their dependencies on external factors such as luminal calcium concentration. In such a situation the model is validated by a possibility of explaining as much experimental data as possible under minimal variation of the model parameters.

RyR channel dynamics includes fast («Franck-Condon») electronic transitions between CP branches, which do not change conformational variable, classical conformational dynamics and classical thermoactivated, and quantum tunneling «non-Franck-Condon» transitions.

One of the ECM features is a key role of quantum effects, in particular, tunneling transitions close to the CP intersection point, where we deal with a degeneracy or quasi-degeneracy. It should be noted that the tunneling transitions do not change conformational variable much and their probability becomes negligibly small with moving off the intersection point. There is no experimental evidence of tunneling effects in RyR channels at present, but tunneling plays key role in many biophysical processes such as electron transfer. The probability of the resonance tunneling between two CP branches was assumed to obey an effective Gamov law:

$$P_{tun} = P_0 e^{-A\,L(Q)\sqrt{\Delta E(Q)}}, \qquad (5)$$

where $L(Q) = \Delta Q_{\pm}$ is the width, and $\Delta E(Q)$ the height of an energetic barrier, or difference between energies in point of tunneling and in the intersection of the CP branches (see Fig. 1), $P_0$ and $A$ are some fitted constants.

Interaction of $Ca^{2+}$ with RyR channel's binding sites is assumed to be of a resonance character, and the probability of the electronic transitions between CP branches was approximated by Lorentzian:

$$P_{elect}(E,\Delta) = \lambda_{elect} \frac{\delta^2}{\delta^2 + (\Delta_{\pm} - E)^2}, \quad (6)$$

where $E$ is the energy of $Ca^{2+}$ ions ($T_{Ca^{2+}} = \dfrac{E}{k_b}$ is an effective "temperature" of $Ca^{2+}$ ions), $\Delta_{\pm} = |E_{+} - E_{-}|$ is the energy of electronic excitation (difference in energies between two CP branches), $\delta$ is a half-width of the resonance peak, $\lambda_{electr} = P_{res} = P_{electr}(E = \Delta)$ is a resonance probability, which is assumed to depend on the $Ca^{2+}$ concentration in a dyadic space.

At present there is no clear understanding of local $Ca^{2+}$ signaling regulation in heart cells, in particular, role of $Ca^{2+}$ ions, which come to dyadic space across sarcolemma through L-type calcium channels ($Ca_{SS}$) and through RyR channels from sarcoplasmic reticulum (SR), and also a role of luminal calcium ($Ca_{jSR}$) is not well understood yet. Unlike purely «electronic» influence of $Ca_{SS}$, influence of pretty slow change of $Ca_{jSR}$ on a RyR channel activation process can be described as a purely «mechanical» one, through conformational stress, which acts on RyR channels [12-14]. Effective stress parameter $p$ is assumed to obey Hill type $Ca_{jSR}$ dependence:

$$p = 2\frac{Ca_{jSR}^n}{Ca_{jSR}^n + K_{Ca}^n} - 1, \quad (7)$$

where $K_{Ca}$ is a value of $Ca_{jSR}$, wherein $p = 0$, $n$ is a parameter, which defines nonlinearity in dependence of $p$ on the luminal $Ca^{2+}$ concentration. Varying $p$ within (-1;1) leads to changes in the RyR channel «energetics» and to a key effect of ECM, that is the change of stable state from the closed to the open one: cC→oO. Typical process of RyR channel opening-closing in terms of the ECM is as follows (Fig.1):

1.   cC → oC: electronic transition from "electronically closed" to "electronically open" branch of CP;

2.   oC → oO: conformational relaxation, in other words, movement on the electronically open branch of CP minima. Channel can move either aperiodically, or damped oscillations near CP minima can occur, depending on the internal friction parameter Γ;

3.   oO* → cO**: tunneling to "electronically closed" branch of CP;

4.   cO** → cO: conformational relaxation to electronically and conformationally closed CP minima.

As it was shown in papers [12-14,22-26] the ECM is able to describe key properties of an isolated RyR channel and RyR clusters in calcium release units. Previous models of calcium dynamics didn't provide an adequate description of RyR channels' opening/closing in $Ca^{2+}$ - release units [27]. Integration of the RyR's ECM into model of calcium dynamics in cardiomyocytes allowed us to simulate and study both spontaneous and stimulated regimes of the calcium release from SR, including calcium oscillatory regimes. In particular, it was shown, that the change of the RyR parameters sensitive to $Ca^{2+}$ concentration both in SR lumen and dyadic space allows us to vary action potential frequency in pacemaker cells in a wide range [26].

### III. CONFORMATIONAL DYNAMICS AND TEMPERATURE EFFECTS IN ECM

Conformational dynamics in ECM is described by Langevin equation:

$$M\ddot{Q} = -\frac{\partial}{\partial Q}E_{\pm}(Q) - \Gamma\dot{Q} + F(t), \qquad (8)$$

where $M$ is an effective mass parameter (in this work M = 1), the first term in the right-hand side describes a total conformational force, $\Gamma$ is effective "internal" friction constant, $F(t)$ is a random force (white noise) [28], where $\overline{F(t)} = 0$ and

$$\overline{F(t)F(t')} = 2\gamma k_B T \delta(t - t'), \qquad (9)$$

where $2\gamma k_B T$ is a fluctuations' spectral density [29], $\delta(t-t')$ is Dirac's $\delta$-function. Random force $F(t)$ is a remaining force acting on a system from an environment in the absence of the systematic friction, $-\Gamma\dot{Q}$ [28]. For a closed system in the thermodynamic equilibrium we deal with the fluctuation-dissipation theorem (FDT) [28], which provides a simple relationship between the system's fluctuations' spectral density and the system's dissipation forces as a consequence of the same nature of system's friction and random fluctuations. Generally speaking, FDT states that Γ=γ.

However, FDT cannot work for our model because of some reasons. We are modeling a nonequilibirum, open, multicomponent and multimodal system. RyR channel is an example of a very complex composite nanoscopic protein, for which the dissipation of conformational dynamics is usually separated into "internal" and "external" ones [30, 31]. The external dissipation is due to an influence of cytoplasm on dynamics of RyR channel's surface atoms and is defined by its viscosity $\eta$: $\gamma \propto \eta$, the internal dissipation is because of an effective protein's viscosity. Unfortunately, there's no reliable data even about cytoplasm's viscosity and about its temperature dependence (see, e.g., [30, 31]). We describe "internal" (Γ) and "external" (γ) friction coefficients as model parameters, assuming an Arrhenius dependence on temperature [31]:

$$\Gamma(T) = \Gamma_0 e^{\frac{E_\Gamma}{k_B T}}, \quad \gamma(T) = \gamma_0 e^{\frac{E_\gamma}{k_B T}}, \qquad (10)$$

with activation energies $E_\Gamma$ and $E_\gamma$ respectively ($k_B$, Boltzmann's constant). Moreover, we will assume that namely the friction coefficients' temperature dependences define temperature effects for RyR-channel, assuming other ECM parameters to be independent of temperature. It's worth noting, that modeling the temperature dependence with energy scale $k_B T$ requires absolute energies for CP branches.

Inequality of internal and external frictions leads to a significant change of temperature dependence of conformational coordinate $Q$ near diabatic CP minima [24]. In the absence of electronic and tunneling transitions conformational variable $Q$ distribution function is follows [24]:

$$\rho(Q) = \frac{1}{\sqrt{2\pi}\sigma} e^{-\frac{(Q-Q_0)^2}{2\sigma^2}} = \frac{1}{\sqrt{2\pi}\sigma} e^{-\frac{E(Q)}{k_B T^*}}, \qquad (11)$$

where $\sigma^2 = \frac{\gamma}{\Gamma} \frac{k_B T}{K}$, $Q_0$ is the conformational variable value in CP branch minima, $E(Q) = \frac{1}{2} K(Q - Q_0)^2$ is elastic energy, $T^* = \frac{\gamma(T)}{\Gamma(T)} T$, effective temperature. When FDT works, Γ=γ, and (11) transforms into well-known "Boltzmann" distribution. It's obvious that effective temperature depends on a real temperature. In case $\Gamma \gg \gamma$ effective temperature is much less than a real temperature that leads to significant temperature effects. Temperature increase leads to population increase of conformational states near CP branches intersection, in other words, to an increase of tunneling transition "open ↔ close" probability of RyR channel. Temperature decrease leads channel "freeze" near CP minima with tunneling suppression accompanied by an increase of time of staying near CP branch's minima.

Mean tunneling transition "open ↔ close" probability can be found as follows [24]:

$$\langle P_{tun} \rangle = P_0 \int_{-\infty}^{+\infty} e^{-A\Delta Q_{\pm}\sqrt{\Delta E}} \frac{1}{\sqrt{2\pi\sigma}} e^{-\frac{E(Q)}{k_B T^*}} dQ \quad . \tag{12}$$

It's obvious, that $\langle P_{tun} \rangle$ rapidly increases close to effective temperature $T_{\pm}^* = \delta E_{\pm}/k_B$, where $\delta E_{\pm}$ is the barrier's height (distance from minima to CP branches intersection), has maxima given $T^* \approx T_{\pm}^*$ and slowly decreases at $T^* > T_{\pm}^*$.

For illustration, in Fig.2 we show temperature dependence of $\langle P_{tun} \rangle$ in a wide temperature range, beyond the physiological range, for transitions "open → close" ($\langle P_{tun} \rangle_{oc}$, solid line) and "close → open" ($\langle P_{tun} \rangle_{co}$, dashed line) when global CP minima is for closed state (p = -1). For typical values of main ECM parameters [16,22-26]: Δ= h = 0, a = 5, K = 10 1 meV energy units were chosen, for other ECM parameters the values were as follows: $P_0$ = 25ms$^{-1}$, $A_{tun}$ = 150, $\Gamma_0$ = 4.3·10$^{-12}$, $E_\Gamma$ = 666 meV (15.2 kcal/mol), $\gamma_0$ =5.6 ·10$^{-5}$, $E_\gamma$ = 51.2 meV (1.2 kcal/mol) [16], which provide quantitative description of experimental data [6]. As it was expected for CP with a closed state in global minima ($\delta E_- > \delta E_+$) a rapid increase of $\langle P_{tun} \rangle_{oc}$ is shifted to lower temperatures compared with $\langle P_{tun} \rangle_{co}$. In physiological temperature range (T~300 K) temperature coefficient Q$_{10}$ is negligibly small for "close → open" transition probability and reaches Q$_{10}$ ~14 for "open → close" transition. In Fig. 2 we also present a temperature dependence of an effective temperature $T^*(T)$. In physiological temperature range (see insert in Fig. 2) $T^*$ is less than 1 K, whereas $T^*=T$ at $T \cong 435$ K.

We see that the features of the tunneling transition between CP branches and the effective temperature concept define the most important properties of temperature effects in ECM of RyR channel.

## IV. THE TEMPERATURE EFFECTS MODELLING OF AN ISOLATED RyR CHANNEL, COMPARISON WITH EXPERIMENTS

Making use of a standard ECM technique [11-13, 22-26], we have performed a series of computer simulations of the RyR gating at different temperatures. Choice of model parameters (see above) was dictated by comparison with experimental data for cardiac sheep RyR channel in a lipid bilayer by Sitsapesan et al. [7] on eye: mean open ($<t_{open}>$) and closed ($<t_{close}>$), open probability $P_{open}$ at 5 and 23 °C, and also a temperature dependence of maximal RyR channel conductance. In addition to conformational dynamics in diabatic CP and tunneling, which are defined by parameters named above, we took into account fast $Ca^{2+}$-activated Frank-Condon electronic transitions cC → oC и oO → cO with probabilities P(cC → oC) = 0.028 ms$^{-1}$, P(oO → cO) = 0.001 ms$^{-1}$, respectively. It's worth noting that in all cases channel's closed state is assumed to be a ground state.

Figure 3 shows a comparison of experimental representative current fluctuations of an isolated RyR channel [7], activated given 10 μM cis-$Ca^{2+}$ concentration and results of the ECM simulations at 5 and 23 °C. Cooling from 23 to 5°C leads to a significant change of the RyR gating by increasing $<t_{open}>$ and decreasing ion current amplitude. These features are primarily related with the increase of the RyR channel's internal friction with a decrease of the temperature.

It's obvious, that ECM allows not only qualitatively but also quantitatively reproduce temperature effects. Potential of the ECM is illustrated in Figs. 4-6, where the temperature dependencies of the RyR channel's $D_{max}$, $P_{open}$, $<t_{open}>$, $<t_{close}>$ are shown. Solid lines are results of modeling; points are experimental data [6]. Dashed, dotted and dash-dotted lines are results of modeling with slightly changed values of parameters $E_\Gamma$, $a$, and P(cC → oC), respectively.

Maximal channel's conductance (Fig.4) at $D_0$=150 and $\beta$=0.65, increases slowly at low temperatures and after reaching a critical temperature ~5 °C rapidly increases with an increase of temperature. It's because of dependence of $Q$(t) on internal friction. At high internal friction (low temperatures) channel moves aperiodically during the conformational relaxation and don't reach high conformational variable values. With an increase of the temperature $T \geq 2$ °C the internal friction becomes less than the critical one $\Gamma \leq \Gamma_{crit} = 2\sqrt{K} = 2\sqrt{10}$ [24] and the channel's conformational dynamics transforms into that of damped oscillations near the electronically open CP branch minima and so channel reaches higher $Q$ values and bigger permittivity $D$.

In Figs. 5-6 the simulated (lines) and experimental (points) temperature dependencies of $P_{open}$, $<t_{open}>$, $<t_{close}>$ are presented. $P_{open}$ has a very special temperature dependence. $P_{open}$ increases until temperature ~10°C so that both $<t_{open}>$ and $<t_{close}>$ also increase. Between ~10°C and ~20 °C, $P_{open}$ and $<t_{open}>$ rapidly fall with $Q_{10}$ ~ 8 and $Q_{10}$ ~ 34, respectively, and $<t_{close}>$ has it's maxima at ~15°C. With further temperature increase $P_{open}$ slowly increases, remaining relatively small, and $<t_{open}>$ slowly comes to zero. $<t_{close}>$ at $T$ >15°C slowly decreases.

Specific "bell shaped" temperature dependencies of kinetic characteristics are explained in terms of the ECM by features of the the tunneling transition frequency temperature dependence. During the conformational relaxation oC→oO channel moves on the electronically open branch across the CP branches intersection, where tunneling probability has its maximal value. At low temperatures and high internal friction there are high tunneling frequencies. With increase of temperature and, as a consequence, decrease of internal friction tunneling frequency decreases and $P_{open}$ increases and reaches its maxima. The further temperature increase transforms channel's conformational dynamics to damped oscillations near CP minima with higher probability of being near CP branches intersection. Frequency of tunneling transitions rises rapidly and hence $P_{open}$ decreases. At high temperatures the channel turns to weakly damped oscillations and the tunneling frequency oO* ↔ cO** in both directions becomes equal, hence $P_{open}$ increases as $<t_{close}>$ decreases.

It's very interesting, that small (4%) decrease of internal friction energy activation $E_Γ$ leads to a significant (~10-15 °C) shift of all curves (dashed lines) on Fig.4-6 to lower temperatures without a significant change of their form. At the same time small (2%) decrease of EC coupling parameter $a$ doesn't significantly change $D_{max}$, but leads to a significant change of $P_{open}$ and $<t_{open}>$ curves'(dotted line) amplitude without significant change both curves' shape and behavior. It's clearly seen, that internal friction and its temperature dependence defines "working temperature range" of RyR channel whereas EC coupling parameter defines the magnitude of temperature effects, in particular, $Q_{10}$ values. Moreover, twofold decrease of electronic transition cC → oC probability to P(cC → oC) = 0.014 мс$^{-1}$ (dash-dotted lines in Fig. 4-6) leads to a small $P_{open}$ amplitude decrease and to other expected results: $<t_{open}>$ doesn't change, and $<t_{close}>$ increased twofold.

## V. CONCLUSION

We have demonstrated that simple electron-conformational model with an assumption of Arrhenius type temperature dependence of "internal and external" frictions, and also a specific "thermo-sensitivity" of an "open ↔ close" tunneling transition, can provide not only qualitative, but also convincing quantitative explanation of the temperature effects on the RyR channel gating. Results of this work open perspective of description of temperature effects on the RyR clusters in $Ca^{2+}$ release units and study modification of $Ca^{2+}$ release from SR by temperature, in particular, temperature dependence of frequency and amplitude of $Ca^{2+}$ sparks.

Unfortunately, these days there is a lack of experimental data about the RyR channel gating modification by temperature that hinders a further development of the model.

Concluding we mark a significant difference between conformational dynamics in the work and that of in the Kramers rate theory [32]. For physiological temperatures an effect can be defined only by internal friction without fluctuations, whereas in the Kramers rate theory fluctuations play a key role.

The research was supported by the Ministry of Education and Science of the Russian Federation, projects #1437 and 2725 (elaboration of the EC model), and the Russian Science Foundation, Project #14-35-00005 (application to the temperature effects in RyR channel functioning).

## REFERENCES

1. D.M. Bers, Nature **415**, 198(2002)
2. R.H. Shutt and S.E. Howlet, Am J Physiol Cell Physiol **295**, C692(2008)
3. H.A. Shiels, M. Vornanen and A.P. Farrell, J Exp Biol **203**, 2771(2000)
4. T.J. Allen and G. Mikala, Pflugers Arch. **436(2)**, 238(1998)
5. T.J. Allen, J Cardiovasc Electrophysiol. **7(4)**, 307(1996)
6. C.L. Elias, X.H. Xue, C.R. Marshall, A. Omelchenko, L.V. Hryshko and G.F. Tibbits, Am J Physiol Cell Physiol **281**, C993(2001)
7. R. Sitsapesan, R.A. Montgomery, K.T. MacLeod and A.J. Williams, J Physiol **434**, 469(1991)
8. D.E. Clapham, Nature **426**, 517 (2003)
9. D.E. Clapham and C. Miller, Proc Natl Acad Sci USA., **108(49)**, 19492(2011)
10. Filip Van Petegem, J Biol Chem. **287(38)**, 31624(2012)

11. S. Györke and C. Carnes, Pharmacol. Ther. **119**, 340 (2008)

12. A.S. Moskvin, M. P. Philipiev, O.E. Solovyova, P. Kohl and V.S. Markhasin, Dokl. Biochem. Biophys. **400**, 32 (2005)

13. A.S. Moskvin, M.P. Philipiev, O.E. Solovyova, P. Kohl and V.S. Markhasin J. Phys. Conf. Ser. 21, **195** (2005)

14. A.S. Moskvin, M.P. Philipiev, O.E. Solovyova, P. Kohl and V.S. Markhasin , Prog. Biophys. Mol. Biol. **90**, 88 (2006)

15. L.A. Irvine, M.S. Jafri and R.L. Winslow, Biophys. J. **76**, 1868 (1999)

16. A. Jara-Oseguera and L. D. Islas, Biophys. J. **104**, 2160 (2013)

17. A. S. Moskvin, B. I. Iaparov, A. M. Ryvkin, O. E. Solovyova, V. S. Markhasin, JETP Letters, **102**, 62 (2015)

18. D. Colquhoun and A. G. Hawkes, Phil. Trans. R. Soc. Lond. B **300**, 1(1982)

19. F. Qin, A. Auerbach and F. Sachs, Proc Biol Sci. **264(1380)**, 375(1997)

20. K. Koshino and T. Ogawa, J. Luminesc. **87–89**, 642 (2000)

21. N. Nagaosa and T. Ogawa, Phys. Rev. B **39**, 4472 (1989)

22. A.S. Moskvin, A.M. Ryvkin, O.E. Solovyova and V.S. Markhasin, JETP Lett. **93**, 403 (2011)

23. A.M. Ryvkin, A. S. Moskvin, O. E. Solovyova and V. S. Markhasin, Dokl. Biol. Sci. **444**, 162 (2012)

24. M.P. Philipiev, Electron-conformational models for calcium release system of the cardiac muscle cells, PhD thesis, Ekaterinburg (2007).

25. A.M. Ryvkin, Study of calcium dynamics in cardiomyocytes based on electron-conformational model of ryanodine-sensitive calcium release channels, PhD thesis, Pushino (2014)

26. A.M. Ryvkin, N.M. Zorin, A.S. Moskvin, O.E. Solovyova, V.S. Markhasin, Biophysics **60** (6), 946 (2015).

27. D. Bers, Excitation-contraction coupling and cardiac contractile force (Springer Science Business Media, 2001),vol. 237, p. 105.

28. W.T. Coffey, Yu.P. Kalmykov, and J.T. Waldron, The Langevin Equation, World Scientific Series in Contemporary Chemical Physics, Third Edition (2012), 852 p.

29. Unlike [16] here we use a common definition of noise term.

30. Ansari, A., Jones, C., Henry, E., Hofrichter, J. & Eaton, W., Science **256**, 1796–1798 (1992).
31. Hagen, S. J., Curr. Prot. Pept. Sci. **11**, 385–395 (2010).
32. H.A. Kramers, Physica (Utrecht) **7**, 284 (1940).

**FIGURES CAPTIONS**

**Fig.1.** Conformational potential of RyR channel in ECM given h=Δ=0 (diabatic potential) with global minima in closed state. Small letters "c,o" are used for electronically closed and open states, respectively, and capital letters "C,O" for conformationally closed and open states, respectively. $\delta E_-, \delta E_+$ - potential barriers for closed and open states, respectively. Vertical arrows point to $Ca^{2+}$-induced Franck–Condon (FC) electronic transitions, horizontal arrow points to a non-FC tunneling transition, downhill arrows point to a conformational dynamics.

**Fig.2.** Temperature dependence of the normalized average tunneling probabilities (12) from electronically open to electronically closed state(solid line) and electronically closed to electronically open(dashed line). Dash-dotted line is a temperature dependence of an effective temperature. Inset shows the 0-40 °C temperature range in a large scale.

**Fig.3.** Representative current fluctuations from a single RyR2 at 23 °C (upper panels) and 5 °C (bottom panels). Left panel: the results of the ECM simulations, right panels: the data taken from[7].

**Fig.4.** Temperature dependence of the RyR maximal conductance: the points are experimental data [7]; curves are the ECM simulation (see text for details)

**Fig.5.** Temperature dependence of the $P_{open}$. The points are experimental data[7], curves are the ECM simulation (see text for details).

**Fig.6.** Temperature dependence of $<t_{open}>$ and $<t_{close}>$: the points are experimental data[7], curves are the ECM simulation (see text for details)

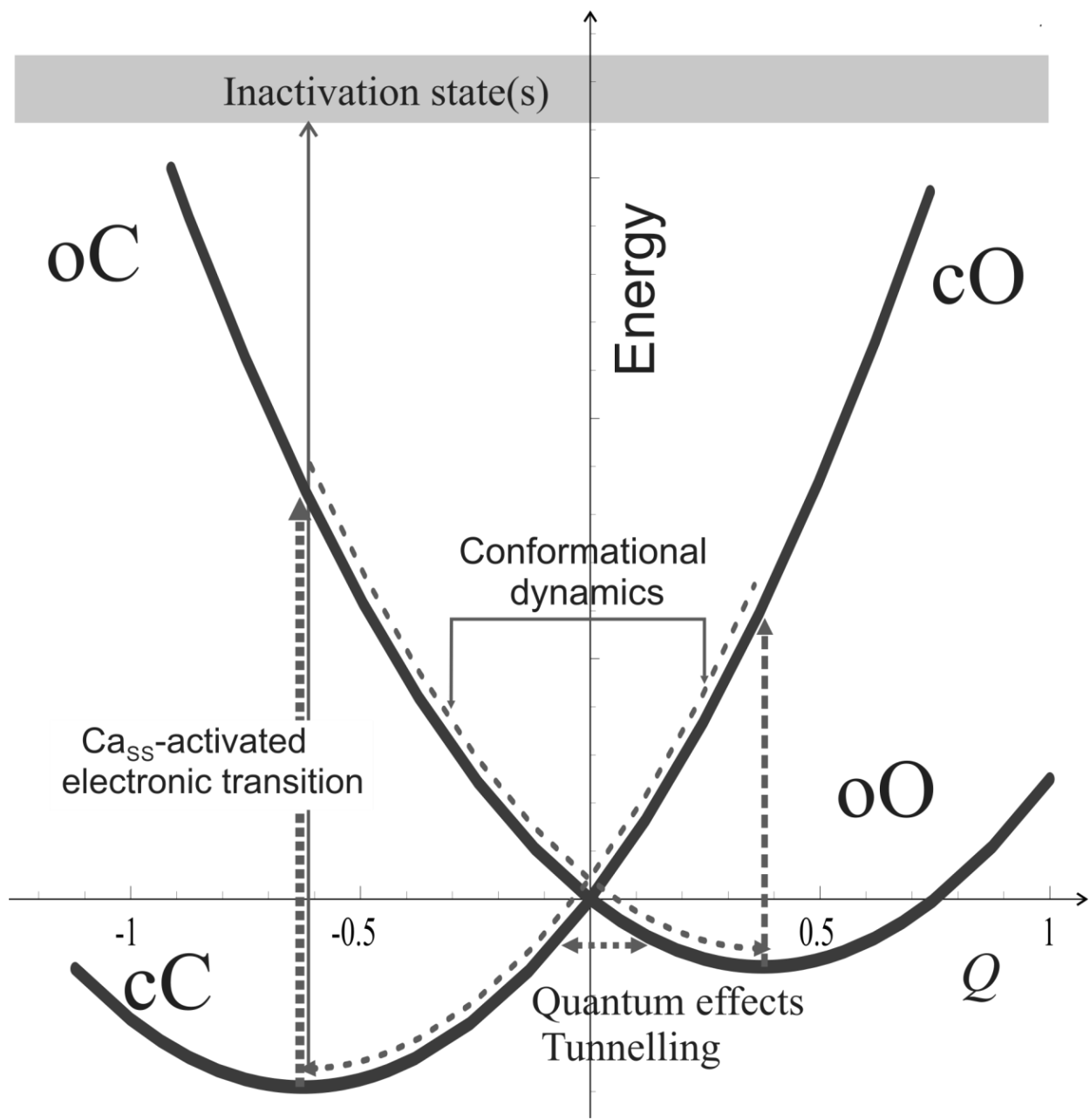

Fig.1

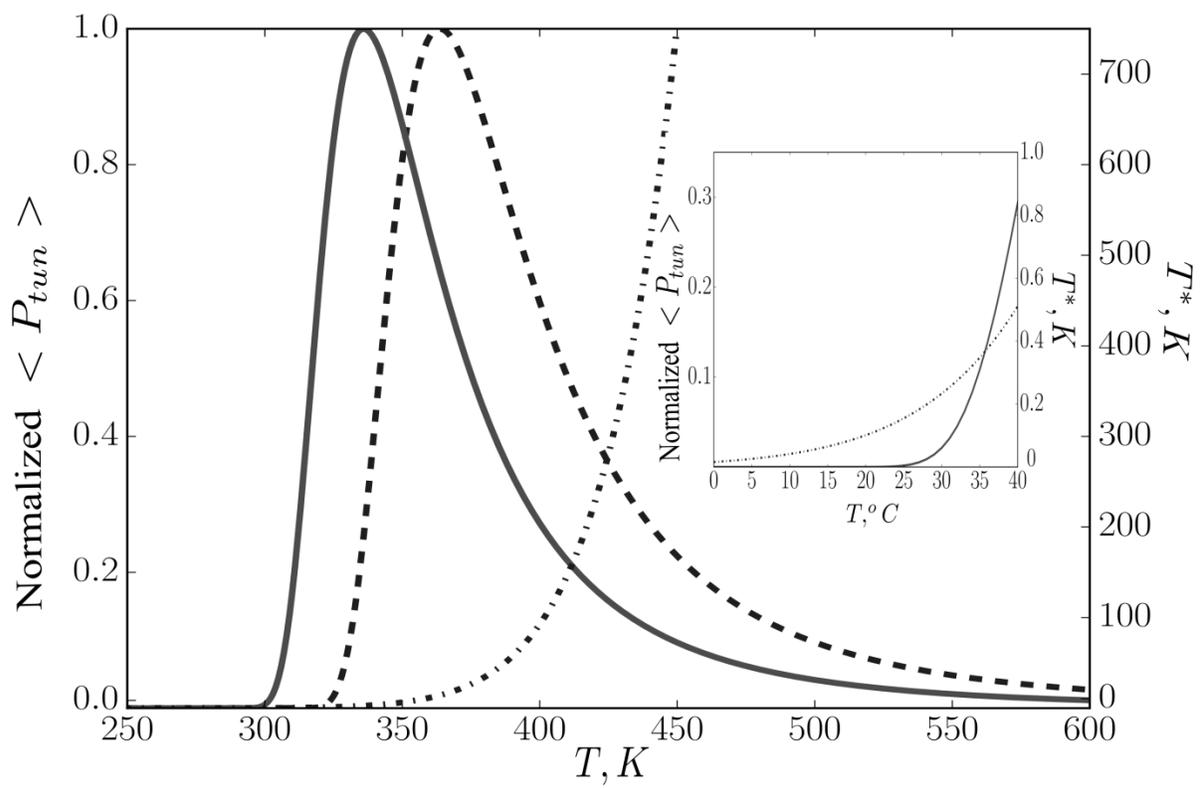

Fig. 2

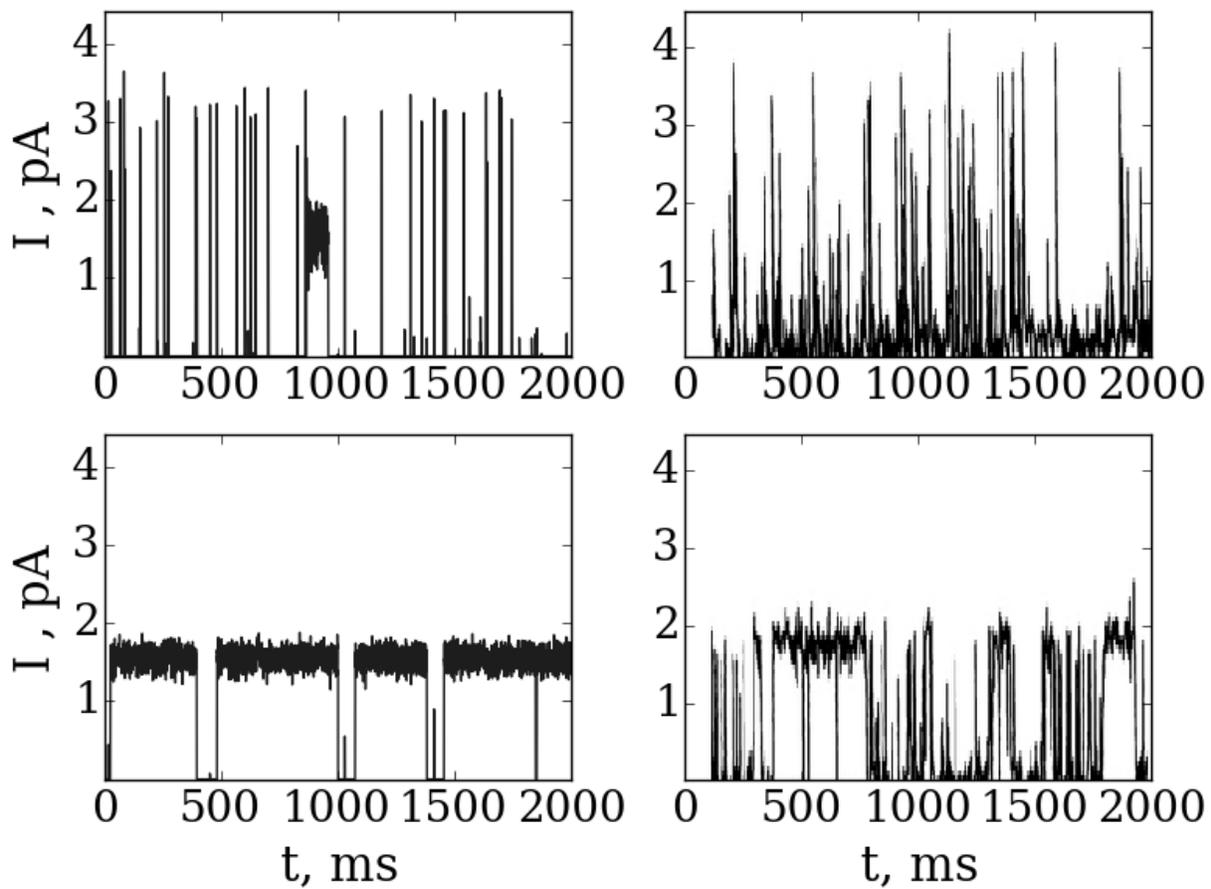

Fig. 3

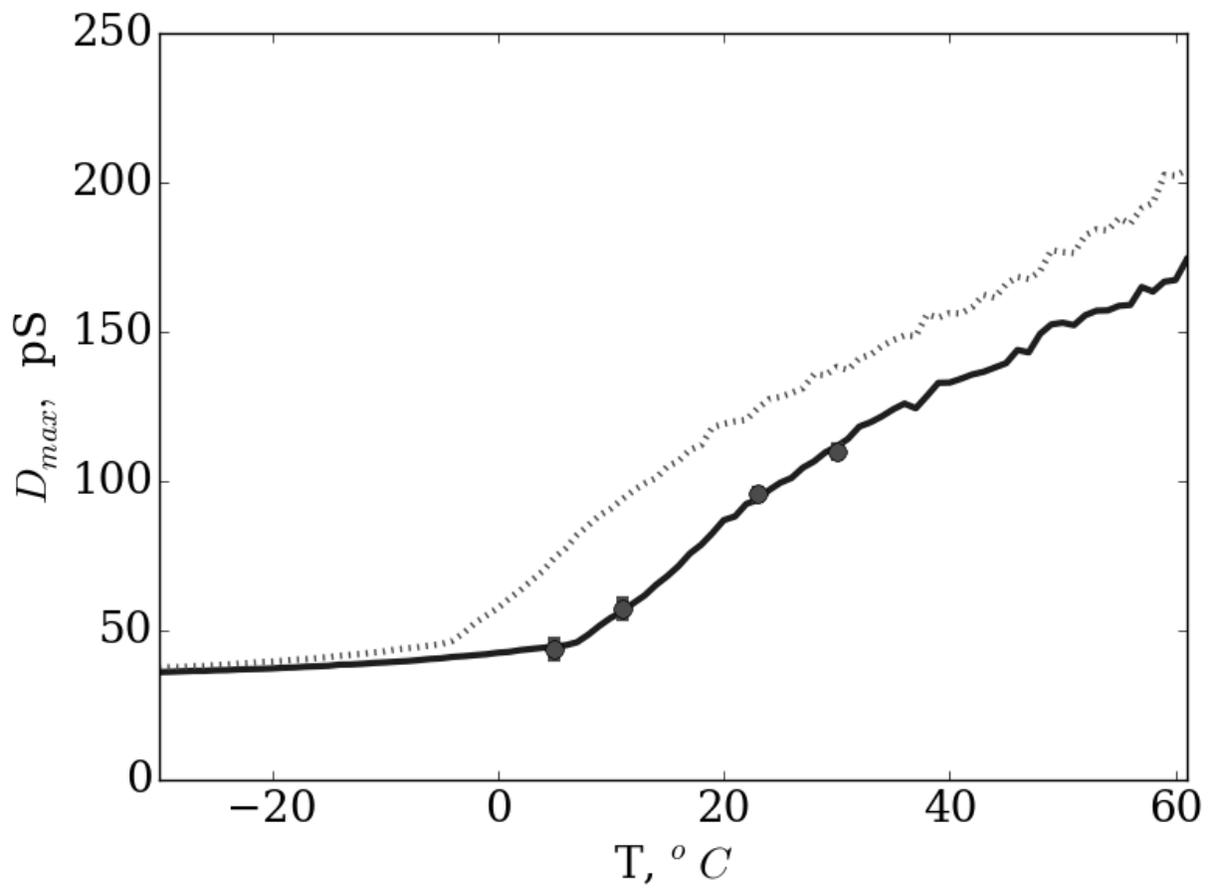

Fig.4

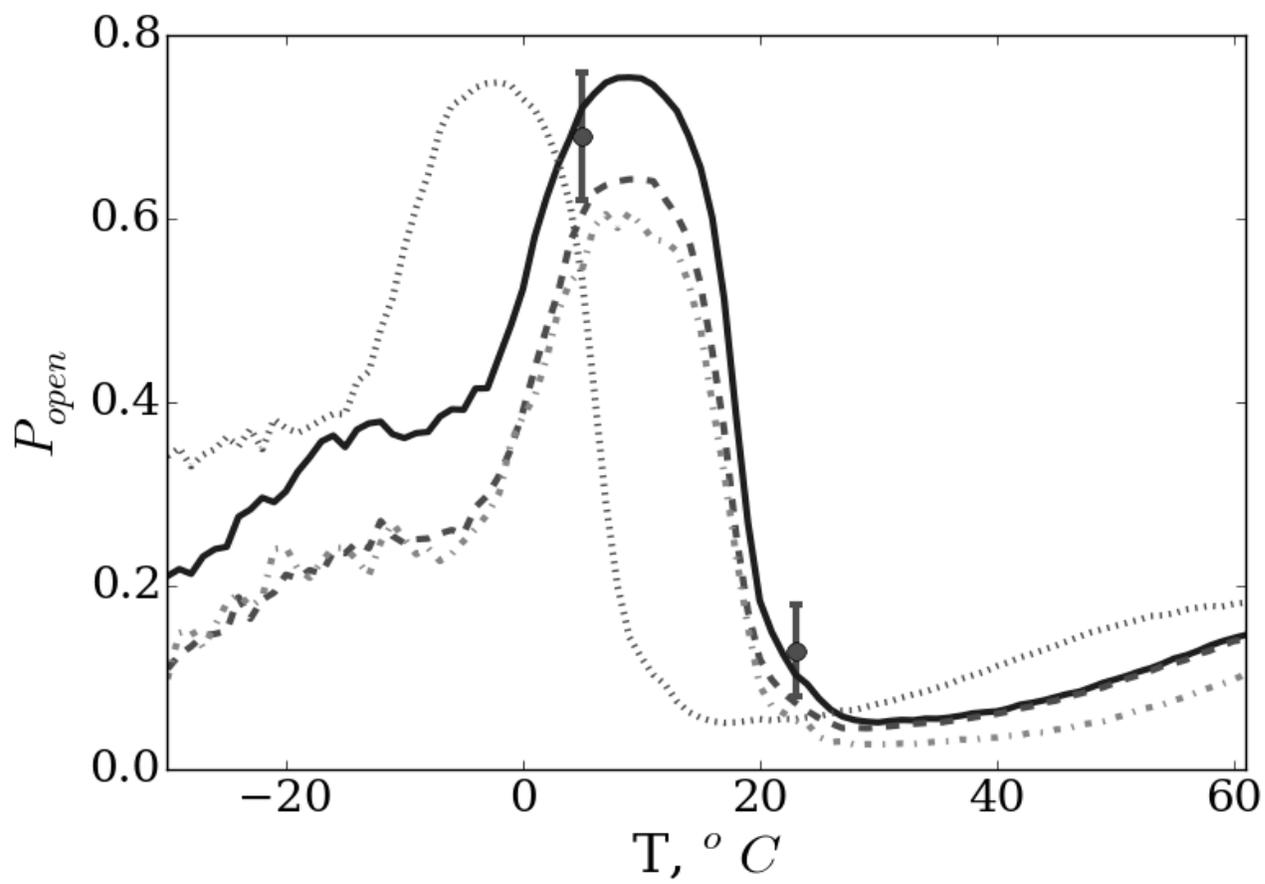

Fig.5

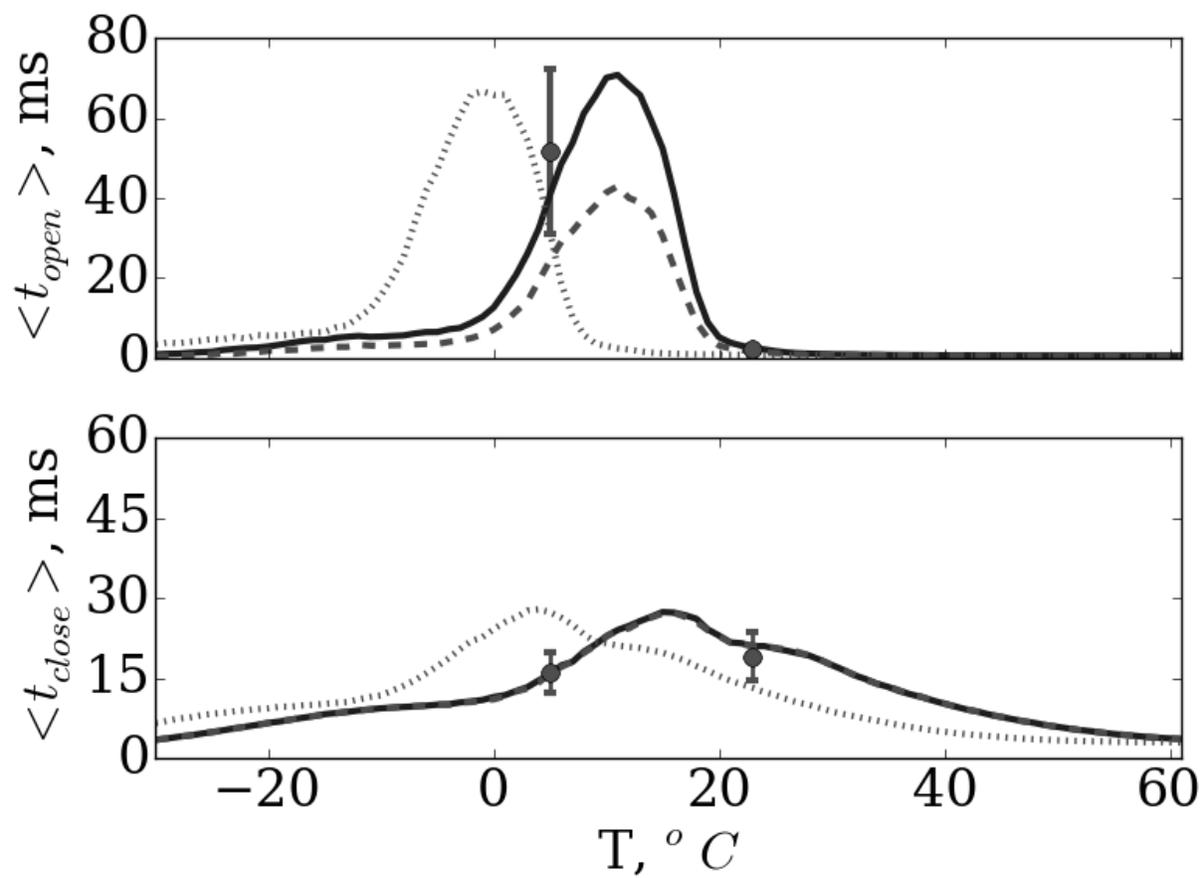

Fig.6